\documentclass[11pt]{article}

\usepackage[a4paper,left=25mm,right=25mm,top=30mm,bottom=30mm]{geometry}
\usepackage{mathpazo}
\usepackage[utf8]{inputenc}
\usepackage[english]{babel}
\usepackage{amsmath,amssymb}
\usepackage{mathrsfs}
\usepackage{multirow}
\usepackage{graphicx}
\usepackage{float}
\usepackage[margin=5mm]{caption}
\usepackage{cite}
\usepackage{color}
\usepackage{sectsty}

\newcommand{\cS}{\mathcal{S}}
\newcommand{\cC}{\mathcal{C}}
\newcommand{\cO}{\mathcal{O}}

\newcommand{\tr}{\mathrm{tr}}
\newcommand{\bx}{\mathbf{x}}
\newcommand{\by}{\mathbf{y}}
\newcommand{\bz}{\mathbf{z}}
\newcommand{\expval}[1]{\langle{#1}\rangle}
\newcommand{\p}{\hphantom{-}}
\newcommand{\m}{-}
\newcommand{\csw}{\textrm{c}_\textrm{sw}}

\definecolor{rossoCP3}{cmyk}{0,0.88,0.77,0.40}

\sectionfont{\fontsize{13}{16}\selectfont}
\subsectionfont{\fontsize{11}{13}\selectfont}

\begin{document}
\begin{center}
 {\bf\Large \color{rossoCP3} Non-perturbative $O(a)$ improvement of the $SU(3)$ sextet model}\\[5mm]
 Martin Hansen\footnote{Electronic address: hansen@cp3.sdu.dk} \\[1mm]
 {\it\footnotesize CP$^3$-Origins, University of Southern Denmark, Campusvej 55, DK-5230 Odense M, Denmark} \\[8mm]
\begin{minipage}{0.85\textwidth}
\hspace{5mm}
We calculate non-perturbatively the coefficient $\csw$ required for $O(a)$ improvement of the SU(3) gauge theory with $N_f=2$ fermions in the two-index symmetric (sextet) representation. For the calculations we impose the standard improvement condition in the Schr\"odinger functional framework, using two different discretisations for the gauge field i.e.~the Wilson plaquette action and the tree-level improved Symanzik (L\"uscher-Weisz) action.
\\[2mm]
{\footnotesize\it Preprint: CP$^3$-Origins-2017-004 DNRF90}
\end{minipage}
\thispagestyle{empty}
\end{center}

\newpage
\section{Introduction}
The $SU(3)$ gauge theory with $N_f=2$ fermions in the two-index symmetric (sextet) representation is believed to be one of the most promising walking Technicolor candidates for a strongly interacting Higgs sector. Perturbative estimates \cite{Dietrich:2006cm} predicts that the model is just below the conformal window, but non-perturbative studies are needed to verify this prediction.

The model has already been investigated extensively by the LatHC collaboration using stout smeared (rooted) staggered fermions and the tree-level improved Symanzik gauge action \cite{Fodor:2012ty}. Their results for the spectrum are consistent with chiral symmetry breaking \cite{Fodor:2012ty,Fodor:2015eea} indicating that the model indeed is below the conformal window. Non-perturbative calculations of the $\beta$-function also does not show any signs of a fixed-point \cite{Fodor:2015zna}.

Similar investigations of the spectrum using unimproved Wilson fermions indicate that the model might be inside the conformal window \cite{Drach:2015sua,Hansen:2016sxp} but the results do not exclude chiral symmetry breaking. While still controversial, calculations using clover improved nHYP smeared Wilson fermions indicates that in fact there might be a fixed-point in the $\beta$-function \cite{Hasenfratz:2015ssa}.

Because cutoff effects can be significant when using unimproved Wilson fermions, further studies with $O(a)$ improvement should provide a more reliable insight into the nature of the model. It should in particular make continuum extrapolations of the spectrum easier.

When applying the Symanzik improvement program for Wilson fermions, it is well-known that $O(a)$ effects can be removed by adding the so-called clover term to the action \cite{Sheikholeslami:1985ij}. However, the coefficient $\csw$ multiplying the clover term must be tuned non-perturbatively to ensure the removal of all $O(a)$ effects.

There is a long tradition for calculating this coefficient in QCD where it has been done for different number of flavors \cite{Luscher:1996ug,Jansen:1998mx,Yamada:2004ja,Tekin:2009kq} and gauge actions \cite{Aoki:2005et,Bulava:2013cta}. In recent years $\csw$ has also been calculated for the SU(2) gauge theory with $N_f=2$ fermions in either the fundamental or the adjoint representation \cite{Karavirta:2011mv}. For the SU(3) sextet model considered here $\csw$ has previously been studied with nHYP smeared fermions \cite{Shamir:2010cq} where it was shown that the tree-level value $\csw=1$ is valid within 20\%. Here we calculate $\csw$ without smearing for two gauge actions i.e. the Wilson plaquette action and the tree-level improved Symanzik action.

The paper is organized as follows. In section~\ref{sec:lattice} we present the lattice action and the improvement condition used to determine $\csw$ non-perturbatively. In section~\ref{sec:results} we discuss the simulations, the data analysis and the results. In section~\ref{sec:conclusion} we conclude and comment on the obtained results.

\section{Lattice formulation\label{sec:lattice}}
Here we start by presenting the simulated lattice action and the improvement condition used to determine $\csw$. The approach used here follows the original papers \cite{Sint:1993un,Luscher:1996vw,Luscher:1996ug} closely.

\subsection{Lattice action}
In the lattice simulations we use the $O(a)$ improved Wilson-Dirac \cite{Wilson:1974sk,Sheikholeslami:1985ij} operator given by
\begin{equation}
 D = D_W + D_{SW} + \delta D,
 \label{eq:D}
\end{equation}
where the first term is the Wilson-Dirac operator, the second term is clover improvement, and the last term is a correction due to the Schr\"odinger functional boundary conditions. The bulk operator is given by
\begin{equation}
 D_W + D_{SW}
 = m_0 + \frac{1}{2}\sum_{\mu}\{\gamma_\mu(\nabla_\mu+\nabla_\mu^*)-\nabla_\mu^*\nabla_\mu\} 
 + \frac{i\csw}{4}\sum_{\mu,\nu}\sigma_{\mu\nu}F_{\mu\nu},
\end{equation}
with $m_0$ being the bare mass and $\csw$ the Sheikholeslami-Wohlert coefficient. The discrete forward and backward covariant derivatives $\nabla_\mu$ and $\nabla_\mu^*$ are given by
\begin{align}
 a\nabla_\mu\psi(x) &= V_\mu(x)\psi(x+a\hat{\mu}) - \psi(x), \\
 a\nabla_\mu^*\psi(x) &= \psi(x) - V_\mu^\dagger(x-a\hat{\mu})\psi(x-a\hat{\mu}),
\end{align}
where $V_\mu(x)$ are the links in the sextet representation as defined in appendix~\ref{sec:appA}. The definition of the Euclidean gamma matrices can be found in \cite{Luscher:1996sc} together with the discretisation of the field strength tensor.

In the Schr\"odinger functional framework we impose periodic boundary conditions along the spatial directions and Dirichlet boundary conditions along the temporal direction \cite{Sint:1993un}:
\begin{equation}
 P_+\psi(0,\bx) = \rho(\bx),\qquad P_-\psi(T-1,\bx) = \rho'(\bx), \qquad P_-\psi(0,\bx) = P_+\psi(T-1,\bx) = 0.
\end{equation}
Similarly, for the conjugate field $\bar{\psi}$ we have
\begin{equation}
 \bar{\psi}(0,\bx)P_- = \bar{\rho}(\bx),\qquad \bar{\psi}(T-1,\bx)P_+ = \bar{\rho}'(\bx), \qquad \bar{\psi}(0,\bx)P_+ = \bar{\psi}(T-1,\bx)P_- = 0.
\end{equation}
In these definitions $P_\pm=\tfrac{1}{2}(1\pm\gamma_0)$ is a projection operator and the $\rho$-fields are source fields for the correlation functions used later on. These source fields are set to zero when generating the configurations, in which case the boundary term in Eq.~\eqref{eq:D} reads
\begin{equation}
 \delta D = (c_F-1)(\delta_{x_0,1}+\delta_{x_0,T-2}).
\end{equation}
This corresponds to a redefinition of the bare mass $m_0$ near the boundaries. For $c_F$ we use the one-loop perturbative value \cite{Luscher:1996vw,Aoki:1998qd} given by
\begin{equation}
 c_F = 1 + c^{(1)}_Fg^2 + \cO(g^4),
\end{equation}
where $c^{(1)}_F$ depends on the gauge action as listed in Table \ref{table:action}.

The gauge action $S_G$ can compactly be written as a sum over all $1\times1$ plaquettes (denoted by $\cS_0$) and $1\times2$ rectangles (denoted by $\cS_1$).
\begin{equation}
 S_G = \beta\sum_{k=0,1}c_k\sum_{\cC\in\cS_k}w_k(\cC)\tr\{1-U(\cC)\}
\end{equation}
The two coefficients $c_k$ are constrained by $c_0+8c_1=1$ to ensure the correct continuum limit. As for the quark fields, we impose periodic boundary conditions along the spatial directions and Dirichlet boundary conditions along the temporal direction.
\begin{align}
 U_k(0,\bx) &= \exp(aC_k),\qquad C_k = \frac{i}{6L}\textrm{diag}(-\pi,0,\pi), \\
 U_k(T-1,\bx) &= \exp(aC_k'),\qquad C_k' = \frac{i}{6L}\textrm{diag}(-5\pi,2\pi,3\pi).
\end{align}
Here $k=\{1,2,3\}$ such that we have spatially uniform boundaries. The weight factors $w_k(\cC)$ are given by
\begin{align}
w_0(\cC) &= 
\begin{cases}
 c_s & \textrm{when $\cC$ is a spatial loop on a boundary} \\
 1   & \textrm{otherwise}
\end{cases} \\
w_1(\cC) &= 
\begin{cases}
 0   & \textrm{when $\cC$ is a spatial loop on a boundary} \\
 c_t & \textrm{when $\cC$ has exactly two links on a boundary} \\
 1   & \textrm{otherwise}
\end{cases}
\end{align}
with
\begin{equation}
 c_s = \frac{1}{2c_0},\qquad c_t = \frac{3}{2}.
\end{equation}
This corresponds to ``Choice B'' in \cite{Aoki:1998qd} and it ensures $O(a)$ improvement at tree-level. This choice of weights is particularly convenient, because the classical minimum of the gauge field is known analytically and it is the same for the two actions considered here i.e. the Wilson plaquette action and the tree-level improved Symanzik action. The coefficients used to specify the two gauge actions can be found in Table~\ref{table:action}.

\begin{table}
\begin{center}
\begin{tabular}{l|ccccc}
          & $c_0$ & $c_1$ & $c_s$ & $c^{(1)}_F$ \\
 \hline\hline\\[-4mm]
 Plaquette & $1$   & $0$   & $\tfrac{1}{2}$ & $-0.0135C_S$ \\[2mm]
 Symanzik & $\tfrac{5}{3}$ & $-\tfrac{1}{12}$ & $\tfrac{3}{10}$ & $-0.0122C_S$ \\[2mm]
 \hline
\end{tabular}
\end{center}
\caption{List of coefficients used in the action. In the last column $C_S$ is the quadratic Casimir for the sextet representation.}
\label{table:action}
\end{table}

\subsection{Improvement condition}
The standard improvement condition \cite{Luscher:1996vw,Luscher:1996ug} relies on the PCAC relation (the axial ward identity) used to define the quark mass on the lattice. In the continuum the relation reads
\begin{equation}
 \langle\partial_\mu A_\mu^a(x)\cO\rangle = 2m\langle P^a(x)\cO\rangle,
\end{equation}
where $A_\mu^a(x)$ and $P^a(x)$ are the axial-vector and pseudoscalar currents, respectively.
\begin{equation}
 A_\mu^a(x) = \bar{\psi}(x)\gamma_\mu\gamma_5\frac{\sigma^a}{2}\psi(x), \qquad
 P^a(x) = \bar{\psi}(x)\gamma_5\frac{\sigma^a}{2}\psi(x).
\end{equation}
On the lattice we define the unrenormalised PCAC relation as
\begin{equation}
 \tfrac{1}{2}(\partial_\mu^*+\partial_\mu)\langle(A_I)_\mu^a(x)\cO\rangle = 2m\langle P^a(x)\cO\rangle,
 \label{eq:PCAClattice}
\end{equation}
where $(A_I)_\mu^a(x)$ is the improved axial-vector current given by
\begin{equation}
 (A_I)_\mu^a(x) = A_\mu^a(x) + c_A\tfrac{1}{2}(\partial_\mu^*+\partial_\mu)P^a(x).
\end{equation}
Here $c_A$ is an additional improvement coefficient. In the unimproved theory, different choices of operators $\cO$ and positions $x$ will generally differ by $O(a)$ effects. For this reason, the improvement condition amounts to requiring that the mass is identical for two different choices of operators and positions, in which case Eq.~\eqref{eq:PCAClattice} holds up to $O(a^2)$ effects.

For the actual calculations we use the two boundary operators $O^a$ and $O'^a$ defined on the lower and upper boundary, respectively.
\begin{align}
 O^a &= \sum_{\by,\bz}\bar{\zeta}(\by)\gamma_5\tau^a\zeta(\bz) \\
 O'^a &= \sum_{\by,\bz}\bar{\zeta}'(\by)\gamma_5\tau^a\zeta'(\bz)
\end{align}
Here the $\zeta$-fields are the boundary quark fields defined from the associated source fields. From the two operators we can define the correlation functions
\begin{align}
 f_P(x_0) &= -\frac{1}{3L^3}\sum_\bx\expval{P_a(x_0,\bx)O^a}, \qquad
 f'_P(T-x_0) = -\frac{1}{3L^3}\sum_\bx\expval{O'^aP_a(x_0,\bx)}, \\
 f_A(x_0) &= -\frac{1}{3L^3}\sum_\bx\expval{A_a(x_0,\bx)O^a}, \qquad
 f'_A(T-x_0) = \frac{1}{3L^3}\sum_\bx\expval{O'^aA_a(x_0,\bx)}.
\end{align}
As suggested in \cite{Luscher:1996sc} the quark mass can now be defined as
\begin{equation}
 M(x_0,y_0) = r(x_0) - \hat{c}_A(y_0)s(x_0),
\end{equation}
where
\begin{align}
 r(x_0) = \frac{(\partial_0^*+\partial_0)f_A(x_0)}{4f_P(x_0)},\qquad
 s(x_0) = \frac{\partial_0^*\partial_0f_P(x_0)}{2f_P(x_0)},\qquad
 \hat{c}_A(x_0) = \frac{r(x_0)-r'(x_0)}{s(x_0)-s'(x_0)}.
\end{align}
Here $\hat{c}_A(x_0)$ is an estimator that allows us to perform the calculations without knowing $c_A$ a priori. We can similarly define the quantity $M'(x_0,y_0)$ by replacing the correlators with their primed counterparts. It now follows that the difference
\begin{equation}
 \Delta M \equiv M(\tfrac{3}{4}T,\tfrac{1}{4}T) - M'(\tfrac{3}{4}T,\tfrac{1}{4}T),
\end{equation}
must vanish up to corrections of $O(a^2)$ when $\csw$ is properly tuned. However, due to the non-trivial background field, there is a small tree-level artefact that must be taken into account to ensure that $\csw=1$ at $g^2=0$. For this reason we require that
\begin{equation}
 \Delta M = \Delta M^{(0)}.
\end{equation}
From measurements on the classical minimum of the gauge field we evaluate the tree-level artefact to be $\Delta M^{(0)}=-0.000651$ on a $8^3\times16$ lattice with bare parameters $m_0=0$ and $\csw=1$. As a consistency check we performed a few simulations at weak coupling to ensure that we correctly reproduce the classical minimum. To complete the definition of the improvement condition we define the quark mass as
\begin{equation}
 M \equiv M(\tfrac{1}{2}T,\tfrac{1}{4}T),
\end{equation}
and impose the improvement condition $\Delta M = \Delta M^{(0)}$ at $M=0$. We could equally well have chosen to define the quark mass via the primed counterpart.

\section{Results\label{sec:results}}
Having properly defined the lattice setup we now turn to the simulations and the obtained results. In connection to the results we want to stress that in this model we naturally expect the value of $\csw$ to be larger than in QCD due to group factors. When comparing the two models, the ratio of quadratic Casimirs $C_S/C_F=2.5$ can be used as a naive estimate for the increase.

\subsection{Simulations}
We have implemented the outlined action and performed a range of simulations with the HMC algorithm \cite{Duane:1987de}. The system is evolved in configuration space by integrating the equations of motion with a multilevel integrator. In most cases we use mass preconditioning \cite{Hasenbusch:2001ne} where the determinant ratio is evaluated on the first level, the HMC term on the second level, and the gauge action on the third level. The second order Omelyan integrator \cite{OMF} has been used in all cases. For the quark determinants we employ even-odd preconditioning \cite{DeGrand:1990dk} to speed up the simulations. Each trajectory has unit integration length and we use an $8^3\times16$ lattice for all simulations.

\begin{table}
\begin{center}
\begin{tabular}{cccllccc}
 $\beta$ &
 $\csw$ &
 $-m_0$ &
 \multicolumn{1}{c}{$M$} &
 \multicolumn{1}{c}{$\Delta M$} &
 MDU &
 $\tau_{int}$ &
 $\csw^\star$ \\
 \hline\hline\\[-4mm]
 \multirow{3}{*}{$30.0$} & $1.0$ & $0.147$ & $\p0.00184(5)$  & $\p0.00108(6)$  &  6000 & $4.5$ & \multirow{3}{*}{$1.115(7)$}\\
                         & $1.1$ & $0.137$ & $\p0.00105(6)$  & $\m0.00058(6)$  &  6000 & $5.6$ \\
                         & $1.2$ & $0.127$ & $\p0.00007(5)$  & $\m0.00180(6)$  &  6000 & $5.5$ \\
 \hline\\[-4mm]
 \multirow{3}{*}{$20.0$} & $1.0$ & $0.235$ & $\m0.00130(7)$  & $\p0.00137(9)$  &  7500 & $6.7$ & \multirow{3}{*}{$1.180(9)$} \\
                         & $1.2$ & $0.200$ & $\p0.00460(7)$  & $\m0.00082(8)$  &  7500 & $6.8$ \\
                         & $1.4$ & $0.170$ & $\p0.00323(9)$  & $\m0.00344(8)$  &  7500 & $6.6$ \\
 \hline\\[-4mm]
 \multirow{3}{*}{$16.0$} & $1.0$ & $0.300$ & $\p0.00192(9)$  & $\p0.00172(11)$ &  7500 & $6.5$ & \multirow{3}{*}{$1.226(7)$} \\
                         & $1.2$ & $0.265$ & $\p0.00254(9)$  & $\m0.00034(11)$ &  7500 & $9.6$ \\
                         & $1.4$ & $0.230$ & $\p0.00158(11)$ & $\m0.00259(11)$ &  7500 & $9.8$ \\
 \hline\\[-4mm]
 \multirow{4}{*}{$12.0$} & $1.0$ & $0.420$ & $\p0.00374(12)$ & $\p0.00240(14)$ & 10000 & $8.5$ & \multirow{4}{*}{$1.351(8)$} \\
                         & $1.2$ & $0.380$ & $\p0.00243(10)$ & $\p0.00047(13)$ & 12500 & $10.2$ \\
                         & $1.4$ & $0.340$ & $\m0.00123(12)$ & $\m0.00097(14)$ & 10000 & $9.9$ \\
                         & $1.6$ & $0.300$ & $\m0.00567(11)$ & $\m0.00274(13)$ & 10000 & $8.6$ \\
 \hline\\[-4mm]
 \multirow{4}{*}{$10.0$} & $1.0$ & $0.522$ & $\p0.00572(12)$ & $\p0.00266(13)$ & 17500 & $9.5$ & \multirow{4}{*}{$1.446(10)$} \\
                         & $1.2$ & $0.480$ & $\p0.00119(12)$ & $\p0.00104(13)$ & 17500 & $11.0$ \\
                         & $1.4$ & $0.435$ & $\m0.00213(12)$ & $\m0.00035(14)$ & 15000 & $10.0$ \\
                         & $1.6$ & $0.380$ & $\p0.00431(11)$ & $\m0.00169(12)$ & 15000 & $10.4$ \\
 \hline\\[-4mm]
 \multirow{4}{*}{$8.0$}  & $1.2$ & $0.640$ & $\p0.00152(13)$ & $\p0.00142(14)$ & 35000 & $10.8$ & \multirow{4}{*}{$1.620(17)$} \\
                         & $1.4$ & $0.590$ & $\m0.00421(13)$ & $\p0.00058(13)$ & 34500 & $11.1$ \\
                         & $1.6$ & $0.530$ & $\p0.00049(12)$ & $\m0.00071(13)$ & 33000 & $11.4$ \\
                         & $1.8$ & $0.470$ & $\p0.00339(13)$ & $\m0.00145(14)$ & 28000 & $10.4$ \\
 \hline\\[-4mm]
 \multirow{4}{*}{$7.0$}  & $1.4$ & $0.708$ & $\m0.00557(13)$ & $\p0.00068(13)$ & 45000 & $9.6$ & \multirow{4}{*}{$1.835(28)$} \\
                         & $1.6$ & $0.645$ & $\m0.00181(12)$ & $\m0.00013(13)$ & 50000 & $10.7$ \\
                         & $1.8$ & $0.580$ & $\p0.00177(11)$ & $\m0.00041(12)$ & 50000 & $10.9$ \\
                         & $2.0$ & $0.513$ & $\p0.00600(12)$ & $\m0.00122(12)$ & 40000 & $11.0$ \\
 \hline\\[-4mm]
 \multirow{4}{*}{$6.5$}  & $1.8$ & $0.650$ & $\p0.00138(13)$ & $\m0.00006(13)$ & 48000 & $9.6$ & \multirow{4}{*}{$2.096(23)$} \\
                         & $2.0$ & $0.586$ & $\m0.00261(14)$ & $\m0.00045(12)$ & 45000 & $9.7$ \\
                         & $2.2$ & $0.518$ & $\m0.00261(13)$ & $\m0.00085(13)$ & 40000 & $11.6$ \\
                         & $2.4$ & $0.450$ & $\m0.00320(13)$ & $\m0.00131(12)$ & 37000 & $11.0$ \\
 \hline\\[-4mm]
 \multirow{4}{*}{$6.0$}  & $1.8$ & $0.736$ & $\m0.00238(15)$ & $\p0.00003(13)$ & 60000 & $10.4$ & \multirow{4}{*}{$2.410(52)$} \\
                         & $2.0$ & $0.665$ & $\m0.00106(13)$ & $\m0.00030(12)$ & 60000 & $10.1$ \\
                         & $2.2$ & $0.590$ & $\p0.00370(13)$ & $\m0.00042(12)$ & 59000 & $10.1$ \\
                         & $2.4$ & $0.517$ & $\p0.00402(12)$ & $\m0.00064(10)$ & 60000 & $10.6$ \\
 \hline
\end{tabular}
\end{center}
\caption{Results for the simulations with the Wilson plaquette action. For each ensemble we quote the measured quark mass $M$ and mass difference $\Delta M$. The integrated molecular dynamics time is given by MDU and $\tau_{int}$ is the integrated autocorrelation averaged across all replicas. In the last column $\csw^\star$ denotes the best fit value to the improvement condition.}
\label{table:plaquette}
\end{table}

\begin{table}
\begin{center}
\begin{tabular}{cccllccc}
 $\beta$ &
 $\csw$ &
 $-m_0$ &
 \multicolumn{1}{c}{$M$} &
 \multicolumn{1}{c}{$\Delta M$} &
 MDU &
 $\tau_{int}$ &
 $\csw^\star$ \\
 \hline\hline\\[-4mm]
 \multirow{3}{*}{$30.0$} & $1.0$ & $0.106$ & $\p0.00112(6)$ & $\p0.00031(7)$ & 3000 & $5.3$ & \multirow{3}{*}{$1.069(4)$} \\
                         & $1.1$ & $0.097$ & $\p0.00064(6)$ & $\m0.00109(6)$ & 3000 & $4.3$ \\
                         & $1.2$ & $0.088$ & $\m0.00015(6)$ & $\m0.00269(7)$ & 3000 & $5.7$ \\
 \hline\\[-4mm]
 \multirow{3}{*}{$20.0$} & $1.0$ & $0.165$ & $\p0.00169(6)$ & $\p0.00078(7)$ & 5000 & $5.8$ & \multirow{3}{*}{$1.110(4)$} \\
                         & $1.1$ & $0.152$ & $\p0.00191(7)$ & $\m0.00056(7)$ & 5000 & $6.4$ \\
                         & $1.2$ & $0.139$ & $\p0.00171(6)$ & $\m0.00174(7)$ & 5000 & $6.1$ \\
 \hline\\[-4mm]
 \multirow{3}{*}{$15.0$} & $1.0$ & $0.227$ & $\p0.00185(8)$ & $\p0.00122(9)$ & 5000 & $7.4$ & \multirow{3}{*}{$1.162(7)$} \\
                         & $1.2$ & $0.196$ & $\p0.00125(8)$ & $\m0.00107(9)$ & 5000 & $7.7$ \\
                         & $1.4$ & $0.166$ & $\m0.00176(8)$ & $\m0.00366(9)$ & 5000 & $7.1$ \\
 \hline\\[-4mm]
 \multirow{4}{*}{$10.0$} & $1.0$ & $0.364$ & $\m0.00227(8)$ & $\p0.00179(10)$ & 12000 & $7.9$ & \multirow{4}{*}{$1.248(4)$} \\
                         & $1.2$ & $0.320$ & $\p0.00123(9)$ & $\m0.00016(9)$  & 12000 & $7.7$ \\
                         & $1.4$ & $0.276$ & $\p0.00250(8)$ & $\m0.00219(10)$ & 10000 & $8.7$ \\
                         & $1.6$ & $0.236$ & $\m0.00206(10)$ & $\m0.00410(11)$ & 8000 & $8.1$ \\
 \hline\\[-4mm]
 \multirow{4}{*}{$8.0$}  & $1.0$ & $0.470$ & $\m0.00115(12)$ & $\p0.00168(13)$ & 12000 & $8.7$ & \multirow{4}{*}{$1.319(8)$} \\
                         & $1.2$ & $0.420$ & $\p0.00213(11)$ & $\p0.00036(13)$ & 12000 & $8.9$ \\
                         & $1.4$ & $0.371$ & $\p0.00233(12)$ & $\m0.00131(13)$ & 10000 & $8.7$ \\
                         & $1.6$ & $0.322$ & $\p0.00116(12)$ & $\m0.00280(12)$ & 10000 & $9.5$ \\
 \hline\\[-4mm]
 \multirow{4}{*}{$6.0$}  & $1.0$ & $0.657$ & $\p0.00150(13)$ & $\p0.00205(13)$ & 30000 & $11.3$ & \multirow{4}{*}{$1.503(14)$} \\
                         & $1.2$ & $0.602$ & $\p0.00176(12)$ & $\p0.00080(12)$ & 30000 & $12.3$ \\
                         & $1.4$ & $0.545$ & $\p0.00178(11)$ & $\m0.00001(12)$ & 30000 & $11.1$ \\
                         & $1.6$ & $0.487$ & $\p0.00132(11)$ & $\m0.00120(12)$ & 25000 & $11.5$ \\
 \hline\\[-4mm]
 \multirow{4}{*}{$5.0$}  & $1.2$ & $0.756$ & $\p0.00215(15)$ & $\p0.00074(14)$ & 40000 & $11.2$ & \multirow{4}{*}{$1.668(26)$} \\
                         & $1.4$ & $0.695$ & $\p0.00093(13)$ & $\p0.00020(13)$ & 40000 & $10.7$ \\
                         & $1.6$ & $0.630$ & $\p0.00233(14)$ & $\m0.00040(13)$ & 35000 & $9.5$ \\
                         & $1.8$ & $0.565$ & $\p0.00160(12)$ & $\m0.00112(13)$ & 35000 & $10.0$ \\
 \hline\\[-4mm]
 \multirow{4}{*}{$4.5$}  & $1.6$ & $0.730$ & $\p0.00257(14)$ & $\m0.00033(13)$ & 50000 & $10.0$ & \multirow{4}{*}{$1.762(30)$} \\
                         & $1.8$ & $0.662$ & $\p0.00109(13)$ & $\m0.00073(13)$ & 50000 & $10.8$ \\
                         & $2.0$ & $0.592$ & $\p0.00030(13)$ & $\m0.00110(13)$ & 45000 & $10.3$ \\
                         & $2.2$ & $0.520$ & $\p0.00080(13)$ & $\m0.00151(13)$ & 40000 & $11.5$ \\
 \hline\\[-4mm]
 \multirow{5}{*}{$3.6$}  & $1.8$ & $0.910$ & $\p0.00169(25)$ & $\m0.00131(19)$ & 57000 & $11.7$ & \multirow{5}{*}{$2.217(38)$} \\
                         & $2.0$ & $0.832$ & $\p0.00026(22)$ & $\m0.00120(19)$ & 57000 & $13.7$ \\
                         & $2.2$ & $0.750$ & $\p0.00255(20)$ & $\m0.00071(17)$ & 57000 & $12.6$ \\
                         & $2.4$ & $0.667$ & $\p0.00331(19)$ & $\m0.00013(17)$ & 47500 & $11.7$ \\
                         & $2.6$ & $0.584$ & $\p0.00140(18)$ & $\m0.00007(17)$ & 38000 & $10.6$ \\
 \hline\\[-4mm]
 \multirow{4}{*}{$3.4$}  & $2.0$ & $0.901$ & $\p0.00721(31)$ & $\m0.00217(21)$ & 60000 & $10.9$ & \multirow{4}{*}{$2.407(25)$}\\
                         & $2.2$ & $0.818$ & $\p0.00439(26)$ & $\m0.00117(19)$ & 60000 & $11.9$ \\
                         & $2.4$ & $0.734$ & $\p0.00092(24)$ & $\m0.00071(18)$ & 50000 & $11.9$ \\
                         & $2.6$ & $0.646$ & $\p0.00270(23)$ & $\m0.00005(19)$ & 40000 & $11.3$\\
 \hline
\end{tabular}
\end{center}
\caption{Results for the simulations with the tree-level improved Symanzik action. For each ensemble we quote the measured quark mass $M$ and mass difference $\Delta M$. The integrated molecular dynamics time is given by MDU and $\tau_{int}$ is the integrated autocorrelation averaged across all replicas. In the last column $\csw^\star$ denotes the best fit value to the improvement condition.}
\label{table:symanzik}
\end{table}

\subsection{Analysis}
For each bare coupling $\beta$ and value of $\csw$ we perform some test runs to determine the bare mass $m_0$ where the quark mass $M$ vanishes. It has previously been shown \cite{Luscher:1996ug,Jansen:1998mx,Tekin:2009kq} that $\Delta M$ only depends weakly on $M$ and for this reason we simply require that $|M| < 0.01$. Once the bare mass has been fixed we perform a full simulation for each set of parameters and for fixed $\beta$ we interpolate linearly in $\csw$ to determine when the improvement condition $\Delta M=\Delta M^{(0)}$ is satisfied. To this end we use the formula
\begin{equation}
 \Delta M-\Delta M^{(0)} = \omega(\csw-\csw^\star),
\end{equation}
where $\csw^\star$ is the value of $\csw$ where the improvement condition is satisfied. The slope $\omega$ depends on the volume, but at fixed volume it is approximately linear in the bare coupling.
\begin{equation}
 \omega = a(1+bg^2)
\end{equation}
The simulations consist of many independent replicas, each containing from 250 to 2000 thermalized configurations. From experience we know that simulations for this model can have long autocorrelations and this is why we choose many independent replicas. To handle autocorrelation within each replica we use the methods in \cite{Wolff:2003sm} to estimate the integrated autocorrelation. The subsequent error analysis is performed by blocked bootstrapping where the block size is chosen as a function of the integrated autocorrelation. After blocking the data we recalculate the integrated autocorrelation to ensure that the blocks are statistically independent.

Finally we want to stress that this model suffers from topological freezing to a much larger extent than QCD. This means that all simulations are carried out in the sector of zero topological charge. This might introduce a small systematic bias in our data, but since the PCAC mass is a UV quantity, it should be relatively insensitive to the topological charge.

\subsection{Plaquette action}
\begin{figure}
 \centering
 \includegraphics[scale=0.94]{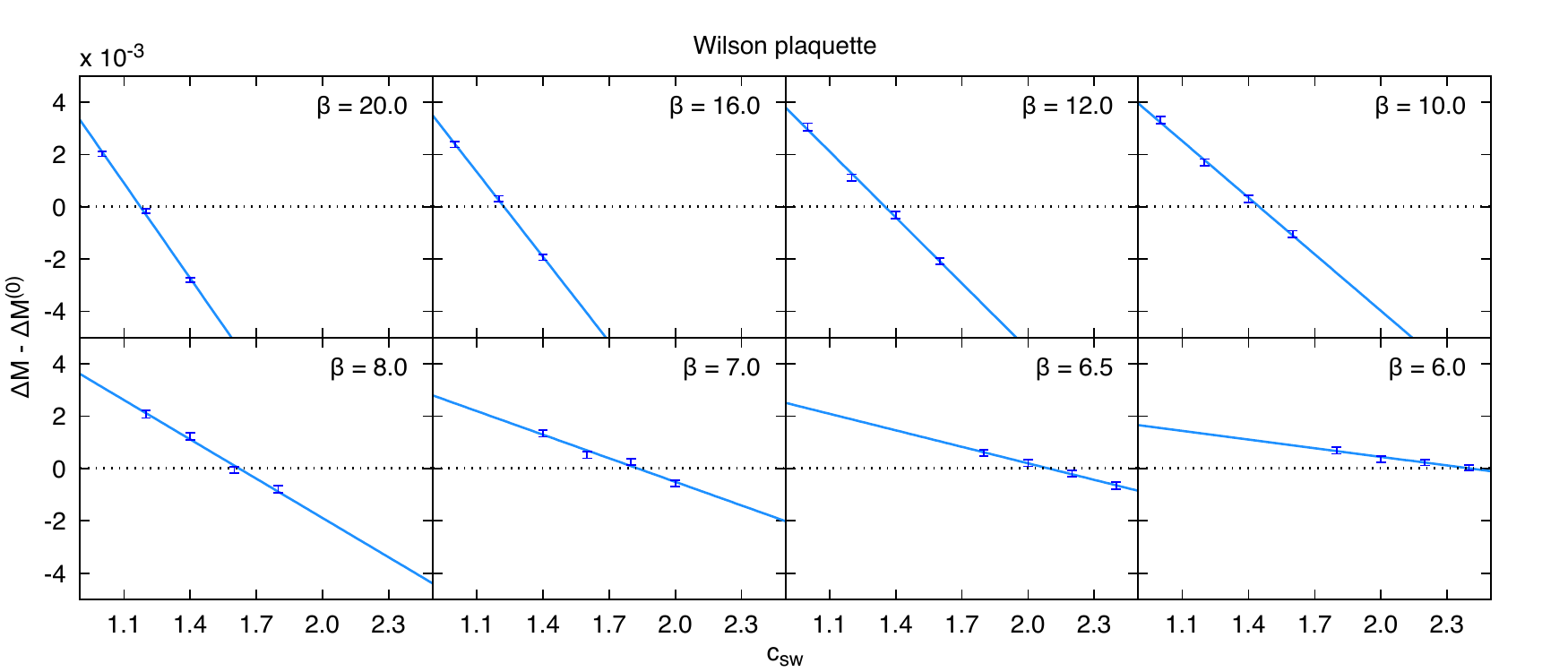}
 \caption{Plot of selected data in Table~\ref{table:plaquette} showing the improvement condition $\Delta M=\Delta M^{(0)}$ for the Wilson plaquette action. We observe that the slope of the observable depends strongly on the bare coupling.}
 \label{fig:dM_wilson}
\end{figure}

\begin{figure}
 \centering
 \includegraphics[scale=1.0]{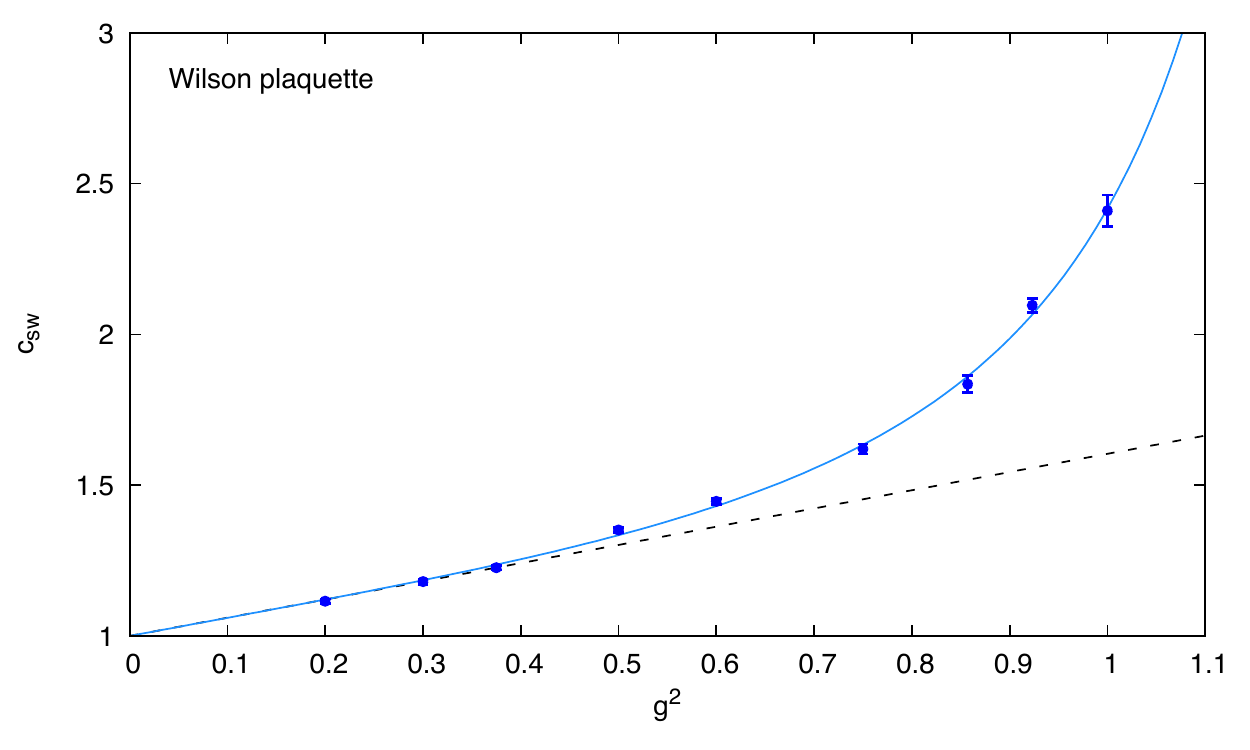}
 \caption{Calculated values for $\csw$ with the Wilson plaquette action. The solid blue line is the interpolating function in Eq.~\eqref{eq:csw_wilson}. The dashed line shows the one-loop result from perturbation theory.}
 \label{fig:csw_wilson}
\end{figure}

The results for the Wilson plaquette action are listed in Table~\ref{table:plaquette} and the improvement condition is plotted in Fig.~\ref{fig:dM_wilson}. In the table we list the bare parameters, the quark mass $M$, the improvement condition $\Delta M$, the total number of trajectories, the integrated autocorrelation, and the best fit value of $\csw$ for each bare coupling. For this action the one-loop perturbative value of $\csw$ was calculated in \cite{Musberg:2013foa} and it reads
\begin{equation}
 \csw = 1 + 0.6039g^2 + \cO(g^4).
 \label{eq:1loopsextet}
\end{equation}
We have chosen to perform several simulations at weak coupling to ensure that we could make connection to perturbation theory. This serves as a stringent test of our entire setup. In Fig.~\ref{fig:csw_wilson} we show the calculated values of $\csw$ together with the interpolating function
\begin{equation}
 \csw = \frac{1-0.1559g^2-0.4578g^4+0.2300g^6}{1-0.7574g^2}.
 \label{eq:csw_wilson}
\end{equation}
The data is well-described by this function in the range $g^2=[0,1]$ with $\chi^2/\textrm{dof}=2.65$. In the perturbative limit, the function correctly reproduces the result in Eq.~\eqref{eq:1loopsextet}.

In QCD the slope of $\Delta M$ depends weakly on the bare coupling, but in this model the dependence is much stronger, as clearly seen on Fig.~\ref{fig:dM_wilson}. From the data we find that the slope is well approximated by the linear function
\begin{equation}
 \omega = -0.0171(1-0.9526g^2).
 \label{eq:slope_plaquette}
\end{equation}
When moving towards stronger coupling, the slope becomes weaker, and ultimately it will change sign. From Eq.~\eqref{eq:slope_plaquette} we can determine that the slope changes sign around $\beta\approx5.7$ and this introduces a problem, because we are unable to determine $\csw$ in the region where the slope is close to zero (in practice when $|\omega|<10^{-3}$). However, we also observe that $\csw$ grows unreasonably large in the region below $\beta=6$ making the action impractical, because it will result in large $O(a^2)$ errors. For this reason we have chosen not to calculate $\csw$ for stronger couplings with this gauge action.

The fact that the slope depends strongly on the bare coupling has another consequence: when the slope is small, the tree-level value $\Delta M^{(0)}$ becomes more important. Taking $\Delta M^{(0)}=0$ will only lead to errors of a few percent in QCD, but in this model the error would be quite substantial.

\subsection{Symanzik action}
\begin{figure}
 \centering
 \includegraphics[scale=0.94]{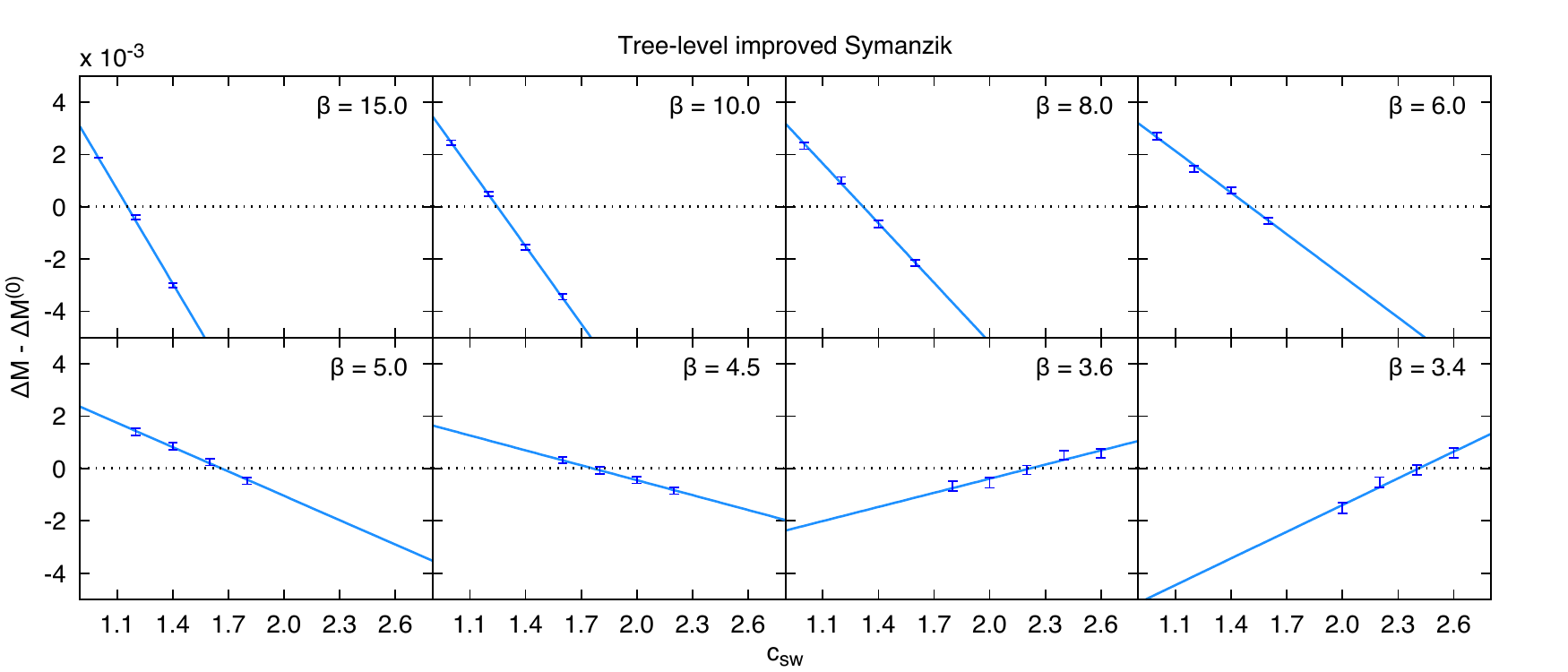}
 \caption{Plot of selected data in Table~\ref{table:symanzik} showing the improvement condition $\Delta M=\Delta M^{(0)}$ for the tree-level improved Symanzik action. At strong coupling we observe that the slope changes sign.}
 \label{fig:dM_symanzik}
\end{figure}

\begin{figure}
 \centering
 \includegraphics[scale=1.0]{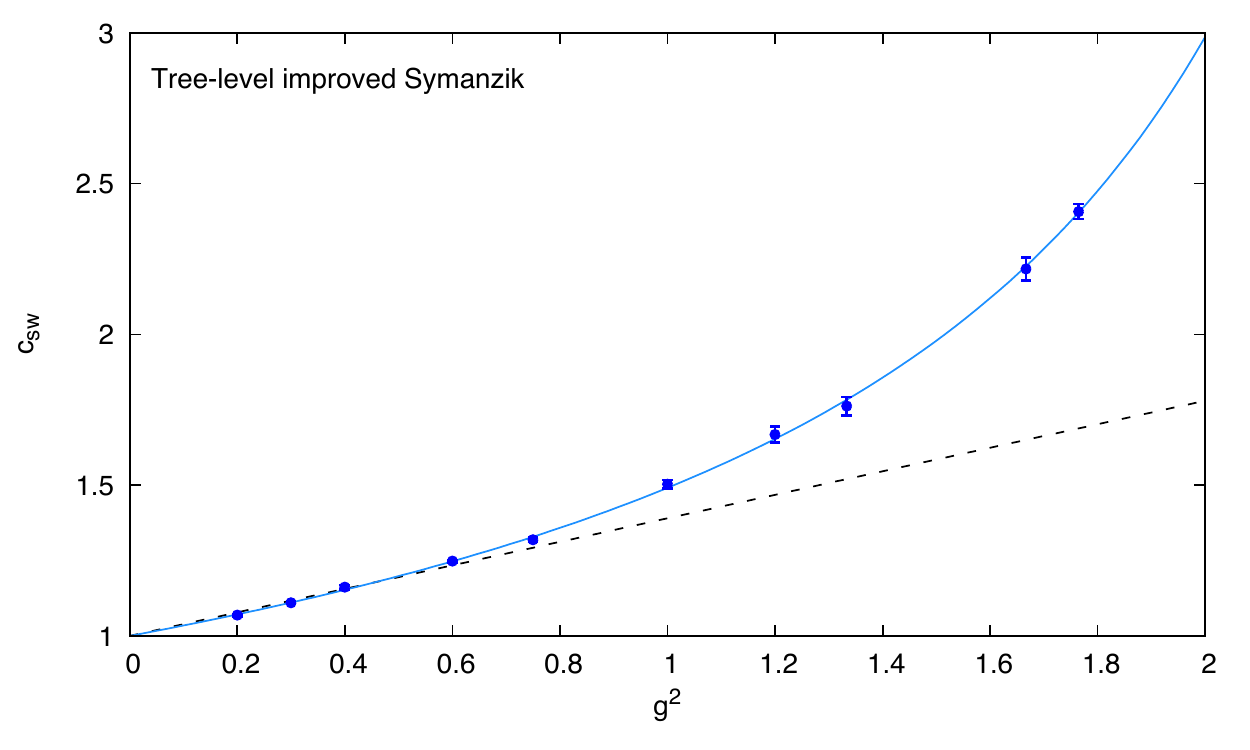}
 \caption{Calculated values for $\csw$ with the tree-level improved Symanzik action. The solid blue line is the interpolating function in Eq.~\eqref{eq:csw_symanzik}. The dashed line shows the central value of our estimated one-loop perturbative result.}
 \label{fig:csw_symanzik}
\end{figure}

To our knowledge no perturbative estimate of $\csw$ exists for the tree-level improved Symanzik action with fermions in higher dimensional representations. However, it is possible to calculate a reasonable estimate from the results in \cite{Aoki:2003sj}. In this paper the authors used conventional (lattice) perturbation theory to calculate $\csw$ for various improved gluon actions and fermions in the fundamental representation. The full calculation requires the evaluation of six Feynman diagrams, but it turns out that the tadpole diagram is the most dominant one, and for this reason they provide an explicit expression.
\begin{align}
 \csw^\mathrm{tadpole} &= \int_{-\pi}^\pi\frac{d^4k}{(2\pi)^4}
 \left\{\left[G_1+G_2\sin^2\left(\frac{k_\nu}{2}\right)\right]\!D_{\mu\mu}(k) + G_3\sin\left(\frac{k_\mu}{2}\right)\sin\left(\frac{k_\nu}{2}\right)\!D_{\mu\nu}(k)\right\}
 \label{eq:int}
\end{align}
Here $D_{\mu\nu}(k)$ is the gluon propagator and $G_i$ are the group constants listed below. We stress that the group constants in the original paper must be rewritten before they are valid for higher dimensional representations. This rewriting consists of the replacement $1/N\to N-2C_R$ after which the group factors read:
\begin{align}
 G_1 &= \tfrac{3}{4}N-\tfrac{1}{2}C_R \\
 G_2 &= 4C_R-\tfrac{3}{2}N \\
 G_3 &= N-4C_R
\end{align}
This rewriting is needed because the identity $1/N=N-2C_F$ is valid only for the fundamental representation\footnote{We are also able to reproduce the result in \cite{Musberg:2013foa} by using the identity to rewrite the result for the plaquette action in \cite{Aoki:2003sj}.}. In the paper it was shown that
\begin{equation}
 \frac{\csw^{(1)}}{\csw^\mathrm{tadpole}} \approx 0.82,
\end{equation}
for both SU(2) and SU(3) with the tree-level Symanzik action. Assuming this ratio is approximately the same for fermions in the two-index symmetric representation, we can estimate the one-loop coefficient $\csw^{(1)}$ for the sextet model by multiplying the integral in Eq.~\eqref{eq:int} with the above ratio.
\begin{equation}
 \csw^{(1)} = 0.82\csw^\mathrm{tadpole} = 0.82\times 0.4734 = 0.39
 \label{eq:csw_symanzik_perturb}
\end{equation}
Here we estimated the integral numerically in a periodic box of size $N^4$ with $N=64$. Using this result, the perturbative estimate reads
\begin{equation}
 \csw = 1 + 0.39g^2 + \cO(g^4).
\end{equation}
Our non-perturbative results for this action can be found in Table~\ref{table:symanzik} and the improvement condition is plotted in Fig.~\ref{fig:dM_symanzik} for at subset of the data. In Fig.~\ref{fig:csw_symanzik} we show the calculated values for $\csw$ together with the interpolating function
\begin{equation}
 \csw = \frac{1-0.0097g^2-0.0129g^4}{1-0.3445g^2}.
 \label{eq:csw_symanzik}
\end{equation}
The data is well-described by this function in the range $g^2=[0,1.8]$ with $\chi^2/\textrm{dof}=0.74$. In the perturbative limit, the interpolating function suggests that our perturbative estimate in Eq.~\eqref{eq:csw_symanzik_perturb} is too large by roughly 15\%. However, as seen on Fig.~\ref{fig:csw_symanzik}, the data does not deviate significantly below $g^2=1$. For this action, the slope of $\Delta M$ is given by
\begin{equation}
 \omega = -0.0165(1-0.6758g^2).
\end{equation}
From this equation we can determine that the slope changes sign around $\beta\approx4.05$ and this is also seen on Fig.~\ref{fig:dM_symanzik} where the two simulations at $\beta=3.6$ and $\beta=3.4$ have opposite slope. Once again this has the unfortunate consequence that we cannot reliably determine $\csw$ in the intermediate region around $\beta\approx4$, but given the smooth behaviour, the interpolating function should work also in this region. In \cite{Shamir:2010cq} it was also observed that the slope of $\Delta M$ had opposite sign at strong coupling.

In Table~\ref{table:mcrit} we show a preliminary estimate of the critical bare mass, i.e.~where the quark mass vanishes, for a few selected values of the bare coupling. By studying the plaquette value in a couple of test simulations we expect a bulk phase transition around $\beta\approx3.2$ and this is in agreement with the location of the bulk phase transition in QCD.

\begin{table}
\begin{center}
\begin{tabular}{ccccc}
 $\beta$ & $\csw$ & $-m_0^\mathrm{critical}$ \\
 \hline\hline\\[-4mm]
 $4.0$ & $1.979$ & $0.721$ \\
 $3.8$ & $2.088$ & $0.734$ \\
 $3.6$ & $2.226$ & $0.741$ \\
 $3.4$ & $2.404$ & $0.732$ \\
 \hline
\end{tabular}
\end{center}
\caption{Preliminary estimate of the critical bare mass  with the tree-level improved Symanzik action. In all cases we expect an uncertainty around $2\times10^{-3}$.}
\label{table:mcrit}
\end{table}

\section{Conclusion\label{sec:conclusion}}
As part of our continued effort to understand the non-perturbative nature of the SU(3) sextet model, we calculated non-perturbatively the coefficient $\csw$ needed for $O(a)$ improved simulations. For the calculations we employed two different discretisations for the gauge action i.e.~Wilson plaquette and tree-level improved Symanzik.

In both cases we observe that the slope of the improvement observable $\Delta M$ depends strongly on the bare coupling, unlike what is observed in QCD, and that it changes sign at strong coupling. This has the unfortunate consequence that we are unable to reliably determine $\csw$ in the intermediate region where the slope changes sign. We speculate that this problem can be solved by choosing another background field, where the slope either changes sign at weaker coupling, making it less problematic, or shows a weaker dependence on the bare coupling altogether.

As expected from group theoretical factors, we also observe that $\csw$ grows large at strong coupling, especially for the Wilson plaquette action. This is a potential problem because it can result in large $O(a^2)$ errors. Fortunately, when using the tree-level Symanzik action, the coefficient is not unreasonably large. The value of $\csw$ can eventually be reduced by resorting to a smeared action, as was shown in \cite{Shamir:2010cq}.

\subsection*{Acknowledgements}
I would like to thank Agostino Patella for hospitality and valuable comments during my stay at CERN, where most of the work was carried out. This work was supported by the Danish National Research Foundation under grant number DNRF90 and by a Lundbeck Foundation Fellowship grant. The computing facilities were provided by the DeIC National HPC Centre at the University of Southern Denmark.

\appendix
\section{Sextet representation\label{sec:appA}}
The represented links $V$ used in the Wilson-Dirac operator can be constructed from the fundamental links $U$ via the map
\begin{equation}
 V^{ab} = \tr(e^aUe^bU^T),
\end{equation}
where $\{e^a\}$ is the orthonormal basis for all real and symmetric $3\times3$ matrices. In our convention for the basis, the represented version of the matrices $C_k$ and $C_k'$ is given by
\begin{align}
 C_k &= \frac{i}{6L}\mathrm{diag}(-2\pi,-\pi,0,0,\pi,2\pi), \\[2mm]
 C_k' &= \frac{i}{6L}\mathrm{diag}(-10\pi,-3\pi,4\pi,-2\pi,5\pi,6\pi).
\end{align}
Furthermore, when using the standard normalization
\begin{equation}
 \tr(T^aT^b) = \tfrac{1}{2}\delta^{ab},
\end{equation}
for the fundamental generators, the quadratic Casimir for the sextet representation reads
\begin{equation}
 C_S = \frac{(N+2)(N-1)}{N} = \frac{10}{3}.
\end{equation}


\begin{thebibliography}{30}
\small

\bibitem{Dietrich:2006cm}
  D.~D.~Dietrich and F.~Sannino,
  \textit{Conformal window of SU(N) gauge theories with fermions in higher dimensional representations},
  Phys.\ Rev.\ D {\bf 75} (2007) 085018
  [hep-ph/0611341].



\bibitem{Fodor:2012ty}
  Z.~Fodor, K.~Holland, J.~Kuti, D.~Nogradi, C.~Schroeder and C.~H.~Wong,
  \textit{Can the nearly conformal sextet gauge model hide the Higgs impostor?},
  Phys.\ Lett.\ B {\bf 718} (2012) 657
  [arXiv:1209.0391 [hep-lat]].



\bibitem{Fodor:2015eea}
  Z.~Fodor, K.~Holland, J.~Kuti, S.~Mondal, D.~Nogradi and C.~H.~Wong,
  \textit{Baryon spectrum in the composite sextet model},
  PoS LATTICE {\bf 2014} (2015) 270
  [arXiv:1501.06607 [hep-lat]].



\bibitem{Fodor:2015zna}
  Z.~Fodor, K.~Holland, J.~Kuti, S.~Mondal, D.~Nogradi and C.~H.~Wong,
 \textit{The running coupling of the minimal sextet composite Higgs model},
  JHEP {\bf 1509} (2015) 039
  [arXiv:1506.06599 [hep-lat]].



\bibitem{Drach:2015sua}
  V.~Drach, M.~Hansen, A.~Hietanen, C.~Pica and F.~Sannino,
  \textit{Conformal symmetry vs. chiral symmetry breaking in the SU(3) sextet model},
  PoS LATTICE {\bf 2015} (2016) 223
  [arXiv:1508.04213 [hep-lat]].



\bibitem{Hansen:2016sxp}
  M.~Hansen and C.~Pica,
  \textit{Sextet Model with Wilson Fermions},
  PoS LATTICE2016 213
  [arXiv:1610.08072 [hep-lat]].



\bibitem{Hasenfratz:2015ssa}
  A.~Hasenfratz, Y.~Liu and C.~Y.~H.~Huang,
  \textit{The renormalization group step scaling function of the 2-flavor SU(3) sextet model},
  arXiv:1507.08260 [hep-lat].



\bibitem{Sheikholeslami:1985ij}
  B.~Sheikholeslami and R.~Wohlert,
  \textit{Improved Continuum Limit Lattice Action for QCD with Wilson Fermions},
  Nucl.\ Phys.\ B {\bf 259} (1985) 572.



\bibitem{Luscher:1996ug}
  M.~Luscher, S.~Sint, R.~Sommer, P.~Weisz and U.~Wolff,
  \textit{Nonperturbative O(a) improvement of lattice QCD},
  Nucl.\ Phys.\ B {\bf 491} (1997) 323
  [hep-lat/9609035].



\bibitem{Jansen:1998mx}
  K.~Jansen {\it et al.} [ALPHA Collaboration],
  \textit{O(a) improvement of lattice QCD with two flavors of Wilson quarks},
  Nucl.\ Phys.\ B {\bf 530} (1998) 185
   Erratum: [Nucl.\ Phys.\ B {\bf 643} (2002) 517]
  [hep-lat/9803017].



\bibitem{Yamada:2004ja}
  N.~Yamada {\it et al.} [JLQCD and CP-PACS Collaborations],
  \textit{Non-perturbative O(a)-improvement of Wilson quark action in three-flavor QCD with plaquette gauge action},
  Phys.\ Rev.\ D {\bf 71} (2005) 054505
  [hep-lat/0406028].



\bibitem{Tekin:2009kq}
  F.~Tekin {\it et al.} [Alpha Collaboration],
  \textit{Symanzik improvement of lattice QCD with four flavors of Wilson quarks},
  Phys.\ Lett.\ B {\bf 683} (2010) 75
  [arXiv:0911.4043 [hep-lat]].



\bibitem{Aoki:2005et}
  S.~Aoki {\it et al.} [CP-PACS and JLQCD Collaborations],
  \textit{Nonperturbative O(a) improvement of the Wilson quark action with the RG-improved gauge action using the Schrodinger functional method}
  Phys.\ Rev.\ D {\bf 73} (2006) 034501
  [hep-lat/0508031].



\bibitem{Bulava:2013cta}
  J.~Bulava and S.~Schaefer,
  \textit{Improvement of $N_f=3$ lattice QCD with Wilson fermions and tree-level improved gauge action},
  Nucl.\ Phys.\ B {\bf 874} (2013) 188
  [arXiv:1304.7093 [hep-lat]].



\bibitem{Karavirta:2011mv}
  T.~Karavirta, A.~Mykkanen, J.~Rantaharju, K.~Rummukainen and K.~Tuominen,
  \textit{Nonperturbative improvement of SU(2) lattice gauge theory with adjoint or fundamental flavors},
  JHEP {\bf 1106} (2011) 061
  [arXiv:1101.0154 [hep-lat]].



\bibitem{Shamir:2010cq}
  Y.~Shamir, B.~Svetitsky and E.~Yurkovsky,
  \textit{Improvement via hypercubic smearing in triplet and sextet QCD},
  Phys.\ Rev.\ D {\bf 83} (2011) 097502
  [arXiv:1012.2819 [hep-lat]].



\bibitem{Sint:1993un}
  S.~Sint,
  \textit{On the Schrodinger functional in QCD},
  Nucl.\ Phys.\ B {\bf 421} (1994) 135
  [hep-lat/9312079].



\bibitem{Luscher:1996vw}
  M.~Luscher and P.~Weisz,
  \textit{O(a) improvement of the axial current in lattice QCD to one loop order of perturbation theory},
  Nucl.\ Phys.\ B {\bf 479} (1996) 429
  [hep-lat/9606016].



\bibitem{Wilson:1974sk}
  K.~G.~Wilson,
  \textit{Confinement of Quarks},
  Phys.\ Rev.\ D {\bf 10} (1974) 2445.



\bibitem{Luscher:1996sc}
  M.~Luscher, S.~Sint, R.~Sommer and P.~Weisz,
  \textit{Chiral symmetry and O(a) improvement in lattice QCD},
  Nucl.\ Phys.\ B {\bf 478} (1996) 365
  [hep-lat/9605038].



\bibitem{Aoki:1998qd}
  S.~Aoki, R.~Frezzotti and P.~Weisz,
  \textit{Computation of the improvement coefficient c(SW) to one loop with improved gluon actions},
  Nucl.\ Phys.\ B {\bf 540} (1999) 501
  [hep-lat/9808007].



\bibitem{Duane:1987de}
  S.~Duane, A.~D.~Kennedy, B.~J.~Pendleton and D.~Roweth,
  \textit{Hybrid Monte Carlo},
  Phys.\ Lett.\ B {\bf 195} (1987) 216.



\bibitem{Hasenbusch:2001ne}
  M.~Hasenbusch,
  \textit{Speeding up the hybrid Monte Carlo algorithm for dynamical fermions},
  Phys.\ Lett.\ B {\bf 519} (2001) 177
  [hep-lat/0107019].



\bibitem{OMF}
  I.~P.~Omelyan, I.~M.~Mryglod and R.~Folk,
  \textit{Symplectic analytically integrable decomposition algorithms: classification, derivation, and application to molecular dynamics, quantum and celestial mechanics simulations},
  Comput.\ Phys.\ Commun.\  {\bf 151} (2003) no. 3 272.


\bibitem{DeGrand:1990dk}
  T.~A.~DeGrand and P.~Rossi,
  \textit{Conditioning Techniques for Dynamical Fermions},
  Comput.\ Phys.\ Commun.\  {\bf 60} (1990) 211.



\bibitem{Wolff:2003sm}
  U.~Wolff [ALPHA Collaboration],
  \textit{Monte Carlo errors with less errors},
  Comput.\ Phys.\ Commun.\  {\bf 156} (2004) 143
   Erratum: [Comput.\ Phys.\ Commun.\  {\bf 176} (2007) 383]
  [hep-lat/0306017].



\bibitem{Musberg:2013foa}
  S.~Musberg, G.~Münster and S.~Piemonte,
  \textit{Perturbative calculation of the clover term for Wilson fermions in any representation of the gauge group SU(N)},
  JHEP {\bf 1305} (2013) 143
  [arXiv:1304.5741 [hep-lat]].



\bibitem{Aoki:2003sj}
  S.~Aoki and Y.~Kuramashi,
 \textit{Determination of the improvement coefficient c(SW) up to one loop order with the conventional perturbation theory},
  Phys.\ Rev.\ D {\bf 68} (2003) 094019
  [hep-lat/0306015].

\end{thebibliography}
\end{document}